\documentstyle[12pt]{article}
\begin{document}

\def\beq{\begin{equation}}
\def\eeq{\end{equation}}
\def\bce{\begin{center}}
\def\ece{\end{center}}
\def\bea{\begin{eqnarray}}
\def\eea{\end{eqnarray}}
\def\ben{\begin{enumerate}}
\def\een{\end{enumerate}}
\def\ul{\underline}
\def\ni{\noindent}
\def\nn{\nonumber}
\def\bs{\bigskip}
\def\ms{\medskip}
\def\wt{\widetilde}
\def\wh{\widehat}
\def\brr{\begin{array}}
\def\err{\end{array}}



\vspace*{3mm}

\begin{center}

{\LARGE \bf
A theory for the conformal factor in quantum R$^2$-gravity}

\vspace{4mm}

\medskip

{\rm {\sc E. Elizalde}
\footnote{E-mail: eli@zeta.ecm.ub.es.
Address june-september 1994:  Department of Physics, Faculty of Science,
 Hiroshima University, Higashi-Hiroshima 724, Japan.
} \\
Center for Advanced Studies, C.S.I.C., Cam\'{\i} de Santa
B\`arbara, 17300
Blanes, \\
and Department E.C.M. and I.F.A.E., Faculty of Physics,
University of  Barcelona, \\ Diagonal 647, 08028 Barcelona,
Catalonia,
Spain

 and

{\sc S.D. Odintsov}
\footnote{E-mail: odintsov@ebubecm1.bitnet.
Also at the Department E.C.M., Faculty of Physics,
University of  Barcelona, Diagonal 647, 08028 Barcelona,
Spain.} \\
Tomsk Pedagogical Institute, 634041 Tomsk, Russia.}

\vspace{5mm}

{\bf Abstract}

\end{center}

A new theory for the conformal factor in R$^2$-gravity is developed.
The infrared phase of this theory, which follows from the
one-loop renormalization group equations for the whole
quantum R$^2$-gravity theory is described. The one-loop effective
potential for the conformal factor is found explicitly and a mechanism
for inducing Einstein gravity at the minimum of the effective
potential for the conformal factor is suggested. A comparison
with the effective theory of the conformal factor induced by the
conformal anomaly,  and also aiming to describe quantum gravity
at large distances, is done.

\vspace{4mm}

\newpage

\ni{\bf 1. Introduction}.
It seems very unlikely nowadays that
the fundamental problems which afflict all theories of the early
Universe ---such as the cosmological constant problem--- may be
solved by simply using some exact symmetry.  Instead, it looks rather
more reasonable the idea that those fundamental problems should be
treated dynamically, on the basis of some approximate symmetries.
In particular, it appears that low-energy (i.e. large distances
or infrared) dynamics of quantum gravity \cite{1,2,3} should be
relevant for
the resolution of the cosmological constant problem.

Some indications in favor of this viewpoint come from the study of
quantum gravity on the De Sitter background \cite{1,4,5,6}, where
it was shown that the graviton propagator may grow without bound in
the infrared (with a dominant contribution from the conformal
part). Then, as it was argued in \cite{2}, the De Sitter space is
not the vacuum of quantum gravity, and conformal factor theory
should be developed as an effective theory for quantum gravity. It
is the purpose of the present letter to develop a theory of the
conformal factor
in quantum $R^2$-gravity in a completely different fashion, as
compared with the anomaly induced model of ref. \cite{2}. We will
show that there exists an infrared asymptotically free solution for
the running coupling constants. Then, the effective potential for
the conformal factor aiming to describe the IR phase of the theory
will be calculated. Applications of the theory to inducing Einstein
gravity with zero cosmological constant will be also discussed.
\ms

\ni{\bf 2. Theory for the conformal factor}.
Let us start from the action of the theory of quantum $R^2$-gravity
\beq
S = \int d^4x \, \sqrt{-g} \left[ \frac{1}{\lambda} W-
\frac{u}{3\lambda} R^2 - \frac{1}{\kappa^2} (R-2\Lambda)\right],
\label{1}
\eeq
where $W = R^2_{\mu\nu} - \frac{1}{3} R^2$, $\lambda$, $u$,
$\kappa^2$ and $\Lambda$ are coupling constants. It is well-known
that a theory with the action (\ref{1}) is multiplicatively
renormalizable \cite{7} and asymptotically free \cite{8} (for a
general review of quantum $R^2$-gravity, see \cite{9}). The only
problem that prevents us from considering the theory (\ref{1}) as
a very serious and consistent candidate for quantum gravity is the
well-known unitarity problem, that so often shows up in the context
of higher-derivative theories. However, it is very possible that
the unitarity problem can be solved here by simply using
non-perturbative methods. But so far only perturbative techniques
(actually, to one-loop) have been developed in quantum
 $R^2$-gravity \cite{9}.

Our first interest will be to discuss the theory (\ref{1}) on
the following background
\beq
g_{\mu\nu} = e^{2\sigma} \eta_{\mu\nu},
\label{2}
\eeq
where $\sigma$ is the so-called conformal factor. Notice that, generally
speaking, $\sigma$ is not a constant field. Substituting
the background (\ref{2}) into the classical action (\ref{1}), one
obtains the classical theory for the conformal factor
\beq
S = \int d^4x \, \left\{ \frac{12u}{\lambda} \left[ \sigma
\Box^2\sigma + 2 (\partial \sigma)^2 \Box \sigma +  (\partial
\sigma)^2 (\partial \sigma)^2\right] - \frac{6e^{2\sigma}
(\partial \sigma)^2}{\kappa^2} + \frac{2\Lambda
e^{4\sigma}}{\kappa^2} \right\}.
\label{3}
\eeq
Here $(\partial \sigma)^2= \eta^{\mu\nu} \partial_\mu \sigma
\partial_\nu \sigma$. Now, starting from this point, one can use
the approximation to the total quantum theory where all spin-two
modes are frozen and only the $\sigma$-field is quantized, and
gives the dominant contribution to the dynamics of the theory.
Then, we immediately see that the infrared phase of quantum gravity
in the sense of ref. \cite{2} is impossible in such a model (see
also the discussion in the next section), just because all the
coefficients of the three higher-derivative terms in (\ref{3}) are
{\it equal}. Hence, we cannot set any of these coefficients equal to
zero, because if we do so any higher-derivative dynamics disappears
and we are left merely with effective theory of the conformal
factor in Einstenian gravity.

Furthermore, there is no need at all to use such
approximation in the above theory. In fact, the $R^2$-gravity
theory (\ref{1}) is multiplicatively renormalizable and the
one-loop $\beta$-functions of this theory are known \cite{8,9}. The
conformal parametrization (\ref{2}) in such a theory is simply a
specific choice of background and all the one-loop calculations can
be carried out in that background, taking into account quantum
corrections coming from all the degrees of freedom. We will show
this below.

First of all, let us write the one-loop renormalization group
equations for the coupling constants of the theory (\ref{1})
\cite{8,9}
\bea
\frac{d\lambda^{-1}(t)}{dt}& =& \frac{133}{10}, \ \ \  \lambda
(0)=\lambda, \nn \\
\frac{du(t)}{dt} &=& -\lambda (t)\left[ \frac{10}{3} u^2(t) +
\frac{183}{10} u(t) +\frac{5}{12} \right], \ \ \ u(0)=u, \nn  \\
\frac{d\kappa^{-2}(t)}{dt} &=& -\frac{\lambda (t)}{\kappa^2(t)}
\left[ \frac{10}{3} u(t) - \frac{13}{6} -\frac{1}{4u(t)} \right],
\ \ \ \kappa^2(0)=\kappa^2, \nn \\
\frac{d\wt{\Lambda}(t)}{dt} &=& \frac{\lambda (t)}{\kappa^4(t)}
\left[ \frac{5}{2} +\frac{1}{8u^2(t)} \right]+  \frac{\lambda
(t)\wt{\Lambda (t)}}{2} \left[ \frac{56}{3} +\frac{2}{3u(t)}
\right], \ \ \ \wt{\Lambda}(0)= \wt{\Lambda},
\label{4}
\eea
where $\wt{\Lambda} \equiv 2\Lambda /\kappa^2$
(the RG equations (\ref{4}) were first correctly obtained
in the harmonic gauge in the paper by Avramidi \cite{8}). Notice that,
as usually,
the last two equations (\ref{4}) are gauge dependent, because here we do
not construct the essential coupling constant as a
combination of $\kappa^2$ and $\Lambda$ \cite{8}.

The solution of the first of the RG equations (\ref{4}) is
\beq
\lambda (t) = \frac{\lambda}{1+ 13.3 \lambda t}.
\label{5}
\eeq
For $\lambda >0$ (what corresponds to the classical positive
action) we get asymptotic freedom in the ultraviolet limit
\cite{8}. In principle, however, considering $R^2$-gravity as an
effective, consistent theory of quantum gravity (let aside the
well-known unitarity problem mentioned above) and hoping that the
$\beta$-functions of quantum $R^2$-gravity (\ref{4}) ---which were
obtained in the ultraviolet domain--- will (at least qualitatively)
describe in fact the infrarred phase of quantum gravity (i.e.,
large distances), we are then allowed to choose the initial value
$\lambda$ to be negative in (\ref{5}),
what leads to a theory that is asymptotically free in the IR limit.
We shall now obtain the IR asymptotic behavior of the solutions of eq.
(\ref{4}). Actually, these equations can be solved explicitly (as
it is easy to see), however the most interesting point for us is
their asymptotic behavior \cite{8}.

Calling $a_1$ and $a_2$ (with $a_1 <a_2$) the roots of the
polynomial in the second of eqs. (\ref{4}), we have
\beq
a_1 = -5.46714, \ \ \ \ \ a_2 = -0.02289,
\eeq
and the solution to this equation can be written as
\beq
u(t) = a_2 + (a_1-a_2) \left[ 1 -
\frac{u-a_1}{u-a_2}\, (1+13.3 \lambda t)^{(a_2-
a_1)/3.99} \right]^{-1}
 \longrightarrow a_2, \ \ \ t \rightarrow  -\infty.
\label{r1}
\eeq
That is, $u(t)$ tends asymptotically to the  fixed point
$a_2$ ---which is the biggest  of the two roots of the second of eqs.
(\ref{4}) (notice that it is still negative).  From the two last eqs.
(\ref{4})), we get the asymptotic behavior
\beq
\kappa^{-2}(t) \sim \frac{\kappa^{-2}}{13.3} \left(- \frac{10}{3}
a_1 + \frac{13}{6} +\frac{1}{4a_1} \right) \frac{1}{1+13.3 \lambda
t}
 \longrightarrow 0, \ \ \ t \rightarrow  -\infty
\eeq
and
\beq
\wt{\Lambda}(t) \sim -\frac{\wt{\Lambda}}{13.3} \left( \frac{28}{3}
+\frac{1}{3a_1} \right) \frac{1}{1+13.3 \lambda t}
 \longrightarrow 0, \ \ \ t \rightarrow  -\infty,
\label{r2}
\eeq
respectively. Again, one may argue that the choice of negative
initial values for $\lambda$ and $u$ is not very convenient for the
action (\ref{1}). Such a choice makes the classical action become
negative at the Euclidean region and this indicates explicitly
the presence of a tachyon. However, if $R^2$-gravity is considered
to be just an effective theory (and not a fundamental one) this may
not be a big drawback and the case deserves consideration. For
the asymptotic behaviors above, we see that we obtain quite
reasonable results, that indicate the existence of an infrared phase
coming from the RG.

Now we are going to consider the effective potential for the
conformal factor (i.e., choosing $\sigma$ to be a constant) ---a
concept that was introduced in refs. \cite{10,11} in a different
context. Notice that this effective potential of the conformal factor
should presumably describe quantum gravity in the infrared phase given
above by the RG equations. Introducing the notation $e^\sigma = \Phi$
and using
the RG improved effective potential (see for example \cite{12}), we get
\beq
V_{RG} (\Phi) = \wt{\Lambda} (t) \Phi^4,
\label{5a}
\eeq
where $\wt{\Lambda} (t) $ is defined as the solution of the RG
equations (\ref{10}) and $t=[2(4\pi)^2]^{-1} \ln (\Phi^2 /\mu^2)$,
$\mu^2 \equiv e^{2\sigma_0}$ (observe that $\mu^2$ is dimensionless).
However, due to the fact that we need a closed, analytic
solution in order to study $V(\Phi)$ (\ref{5a}) ---and that we get
explicitly only  approximate asymptotic solutions--- in our
specific analysis we will have to restrict ourselves to a one-loop
(non-improved) potential only. This can be easily obtained from
(\ref{5a}). For example, using the Coleman-Weinberg normalization
conditions, it becomes
\bea
V^{(1)} (\Phi)&=& \wt{\Lambda}  \Phi^4 + \frac{1}{2(4\pi)^2} \left[
 \frac{\lambda}{\kappa^4} \left( \frac{5}{2} + \frac{1}{8u^2}
\right) +  \lambda\wt{\Lambda} \left( \frac{28}{3} + \frac{1}{3u}
\right) \right]  \Phi^4 \left( \ln \frac{\Phi^2}{\mu^2}- \frac{25}{6}
\right) \nn \\
 & \equiv &  \wt{\Lambda}  \Phi^4 + A  \Phi^4  \left( \ln
\frac{\Phi^2}{\mu^2}- \frac{25}{6} \right).
\label{6}
\eea

In the above effective potential for the conformal factor, {\it
all} degrees of freedom of the consistent theory of $R^2$-gravity
are taken into account, and not just the ones induced by the
quantum conformal mode, as was done in \cite{10}, or either just part of
the contribution from $R^2$-gravity, as was done in \cite{11}.
It is not difficult to analyze the one-loop effective potential
(\ref{6}) for different choices of the coupling constants. The
v.e.v. for the conformal factor is given by
\beq
\frac{\sigma_{vev}}{\sigma_0} = \frac{11}{3}-
\frac{\wt{\Lambda}}{A}, \ \ \ \ \ A>0,
\label{7}
\eeq
which depends, of  course, on the choice of the parameters of the
theory. Let us recall the fact that the potential is defined only
on the positive half-axis of $\Phi$. Further, we observe that the
classical (as well as the one-loop) potential is defined for the
value $\Phi =0$ only asymptotically (since this is obtained for
$\sigma \rightarrow -\infty $). This means that the classical
potential grows monotonically and yields no preferred value for the
conformal factor. Such a value, given by (\ref{7}), appears only as
a result of quantum corrections.

One can extend the above picture to a more dynamical situation, by
choosing the background metric in the following form:
\beq
g_{\mu\nu} = \wh{g}_{\mu\nu} e^{2\sigma},
\label{8}
\eeq
where $ \wh{g}_{\mu\nu}$ is some curved metric (with constant
curvature) and the conformal factor $\sigma$ is a constant one.
Using then the technique of ref. \cite{13} and the explicit form of
the $\beta$-functions, one can find the one-loop effective potential
for the conformal factor as follows (in the linear
curvature approximation and using the normalization precriptions of
ref. \cite{13})
\bea
V^{(1)} (\Phi)&=& \wt{\Lambda} \Phi^4- \frac{1}{\kappa^2}\wh{R}
\Phi^2  + \frac{\wh{R} \Phi^2 }{2(4\pi)^2}
\frac{\lambda}{\kappa^2} \left( \frac{10}{3}u - \frac{13}{6} -
\frac{1}{4u} \right) \left( \ln \frac{\Phi^2}{\mu^2}-3 \right) \nn
\\ &&
+ \frac{\Phi^4 }{2(4\pi)^2} \left[ \frac{\lambda}{\kappa^4} \left(
\frac{5}{2} + \frac{1}{8u^2} \right) +  \lambda\wt{\Lambda} \left(
\frac{28}{3} + \frac{1}{3u} \right) \right]  \left( \ln
\frac{\Phi^2}{\mu^2}- \frac{25}{6} \right) \nn \\
 & \equiv &  \wt{\Lambda}  \Phi^4- \frac{1}{\kappa^2}\wh{R}
\Phi^2  + A  \Phi^4  \left( \ln \frac{\Phi^2}{\mu^2}- \frac{25}{6}
\right)- B\wh{R} \Phi^2  \left( \ln \frac{\Phi^2}{\mu^2}-3 \right).
\label{9}
\eea
Here we suppose that the conformal factor in the metric (\ref{8})
gives the dominant contribution to the effective potential. Such
hypothesis
is very reasonable for the early universe (before the GUT epoch) in an
infrared phase for quantum gravity of the type (\ref{r1})-(\ref{r2}).
Then,
in the effective action for quantum $R^2$-gravity it is reasonable to
expand this action in powers of the constant curvature $\wh{R}$, as
 was done above.

We are now ready to consider the conformal factor as being dynamical, in
the same frame of an approximate description of quantum gravity as
an effective theory. Then, there appears the very interesting
possibility of inducing Einstein gravity in the infrared phase, as
a result of a first-order $\wh{R}$-curvature induced phase
transition in the effective potential for the conformal factor.

 Indeed, let us make the potential (\ref{9}) be dimensionless, by
considering, as usually, $\kappa^4 V /\mu^4$ and by further
introducing the dimensionless variables
\beq
x \equiv \frac{\Phi^2}{\mu^2}, \ \ \ y \equiv \frac{|\wh{R}|
\kappa^2}{\mu^2}, \ \ \ \epsilon = \mbox{sgn} \ \wh{R},
\label{10}
\eeq
where $\mu^2 =e^{2\sigma_0}$, as above. The standard conditions for
a first-order curvature-induced phase transition are \cite{14} (a
first example of a curvature-induced phase transition in QED in
curved space has been given in \cite{14})
\beq
V(x_c,y_c)=0, \ \ \ \ \left. \frac{\partial V}{\partial x}
\right|_{x_c,y_c} =0, \ \ \ \ \left. \frac{\partial^2 V}{\partial
x^2} \right|_{x_c,y_c} >0.
\label{11}
\eeq
After the phase transition has taken place, the effective action
in the minimum has the form of classical Einstein gravity
\beq
\Gamma = - \frac{1}{16 \pi G_{ind}} \int d^4x \, \sqrt{- \wh{g}}
\left( \wh{R} - 2 \Lambda_{ind} \right),
\label{12}
\eeq
where the general expressions for $\Lambda_{ind}$ and $G_{ind}$ may
be found in analogy with ref. \cite{13} (see also \cite{9}).
Instead of doing this, we will present some examples which show the
possiblity of inducing (\ref{12}).

Let us choose $u<<1$, $1/u > \wt{\Lambda}  \kappa^4$, $\lambda /u
< 10^3$, $ \wt{\Lambda}  \kappa^4 <<1$, and $\lambda <<1$. Then, by
analysing eqs. (\ref{11}) we get
\beq
\varphi_c^2 \simeq e^{19/6} \mu^2 \ (\sigma_c \simeq 19/12 +
\sigma_0), \ \ \ \ \wh{R}_c \simeq -e^{19/6} \frac{\epsilon \lambda
\mu^2}{16(4\pi)^2 u^2\kappa^2}.
\label{13}
\eeq
In particular, for $u \simeq 10^{-2}$ and $ \lambda \simeq 10^{-4}$
we obtain a reasonable estimation that does not violate our
approximation, what proves that the phase transition is possible.
Then,
\bea
\frac{1}{16 \pi G_{ind}} &\simeq & \frac{\varphi_c^2}{\kappa^2}
\left( 1 + \frac{\lambda}{48 (4\pi)^2 u} \right), \nn \\
\frac{\Lambda_{ind}}{8 \pi G_{ind}} &\simeq & \varphi_c^4 \left(
\wt{\Lambda} - \frac{\lambda}{16 (4\pi)^2 u^2 \kappa^4}
\right).
\label{14}
\eea
As is easy to see, after convenient fine-tuning of the parameters
of the theory (always within the allowed range of our
approximation) we can get indeed $\Lambda_{ind} =0$. In principle,
one could suggest other choices for the parameters of the theory, for
which the mechanism described above to induce Einstein
gravity could also work. Of course, taking this mechanism as a
serious possibility, we ought eventually to apply the point of view
of ref. \cite{11}, where the metric $\wh{g}_{\mu\nu}$ was
considered to be the physical metric corresponding to the real
low-energy world.
\ms

\ni{\bf 3. Anomaly induced theory for the conformal factor}.
Let us now compare our theory with the effective theory of the
conformal factor suggested in ref. \cite{2}. Starting from the
conformal anomaly  \cite{15} (for a general discussion see
\cite{16})
\beq
T_\mu^{\ \mu} = b\left( F + \frac{2}{3} \Box R \right) + b' G + b''
\Box R,
\label{15}
\eeq
where the contributions of the fields of spin $0, 1/2,$ and $1$ to
the coefficients of  (\ref{15}) are known, one can integrate over
the conformal anomaly and thus get an anomaly-induced action
\cite{17}. Choosing, for simplicity, the conformal parametrization
as in (\ref{2}), we get the following \cite{2,17}
\beq
S_0 = \int d^4x \, \left\{ - \frac{Q^2}{(4\pi)^2} (\Box \sigma)^2-
\zeta \left[ 2 (\partial \sigma )^2 \Box \sigma +(\partial \sigma
)^2 (\partial \sigma )^2  \right] - \frac{6e^{2\sigma}}{\kappa^2}
(\partial \sigma )^2 +  \frac{2\Lambda e^{4\sigma}}{\kappa^2}
\right\},
\label{16}
\eeq
where $Q^2/(4\pi)^2 =2b +3b'$ and $\zeta =2b+2b'+3b''$.  As in
expression (\ref{3}), the last two terms give the contribution of
the classical Einstenian action.

It was suggested in \cite{2} that spin-2 gravitational modes do not
change the structure of the action that was considered as an
effective action for quantum gravity in the infrared phase ($\zeta
=0$). This proposal may be justifiable partially by the fact that
inclusion of spin-2 gravitational modes in different models may be
done by means of finite renormalization of the conformal anomaly
coefficients. In other words, spin-2 quantum gravity modes give an
additional contribution to $T_\mu^{\ \mu}$ \cite{18,19}.

However, as one can see if one starts from $R^2$-gravity without
matter, the infrared phase in the sense of paper \cite{2} {\it
cannot} be realized, just because in (\ref{3}) all the coefficients
of the higher-derivative terms are {\it equal}. Of course, if one
adds to the system conformal invariant matter, then the
anomaly-induced action corresponding to matter gives a finite
renormalization of the coefficients in (\ref{3}) (working in the
spirit of \cite{2}), and then the IR phase may be realized for
$\zeta=0$. But in this case we actually loose  the direct
interpretation (as a background) developed above, and are obliged to
invent some
kind of synthesis of the two approaches (what looks rather artificial).
\ms

\ni{\bf 4. Concluding remarks}.
In this letter we have discussed a consistent theory of the
conformal factor in quantum $R^2$-gravity. In particular, we have
calculated the effective potential for the conformal factor at the
one-loop level and we  have shown how it can be applied to
inducing Einstein gravity. We have compared our theory with an
alternative one in which the conformal factor is anomaly induced.
It is remarkable that through our simple formulation the coupling
of the conformal sector with interacting matter can be so naturally
discussed. For, notice that if one would try to do the same using
the anomaly-induced conformal factor theory, one would readily fall
into difficulties, owing to the fact that the matter interaction
terms (as $\lambda \varphi^4$ itself) induce an additional
$R^2$-term in the conformal anomaly at the multiloop level.  This
term breaks the consistency conditions and, also, it cannot be
simply integrated in the conformal anomaly in a unique way ---in
order to give the corresponding part in the anomaly-induced action
\cite{17}.

In order to see this point with an explicit example, let us consider the
 $\lambda \varphi^4$-theory with the action
\beq
S_m = \int d^4x  \, \sqrt{-g} \left( \frac{1}{2} g^{\mu\nu}
\partial_\mu \varphi \partial_\nu \varphi+  \frac{1}{2} \xi R
\varphi^2-   \frac{1}{4!} f \varphi^4 \right),
\label{17}
\eeq
interacting with the theory (\ref{1}), where for simplicity we set
$\kappa^{-2} =\Lambda =0$ (no dimensional coupling constants). Such
theory is multiplicatively renormalizable (see the last of refs.
\cite{8} and \cite{9}). Working in the background metric (\ref{8})
and in the linear curvature approximation (i.e., $\varphi^2 >>
\wh{R}$, $\sigma =$ const.), and using the results of refs. \cite{8}
(last reference) and \cite{20} (where the one-loop $\beta$-functions
have been given explicitly), one finds
\bea
V(\varphi ) &=& - \frac{1}{2} \xi e^{2\sigma} \wh{R} \varphi^2 +
\frac{1}{4!} f e^{4\sigma} \varphi^4 + \frac{1}{32(4\pi)^2} \left\{
3f^2+\lambda^2\xi^2 \left[ 15+ \frac{3}{4u^2} + \frac{9\xi}{u^2}
( 3\xi -1) \right] \right. \nn \\
&&+ \left. \lambda f \left[ -\frac{28}{3} + 8 \xi + \frac{1}{u}
\left( -18 \xi^2 + 8 \xi - \frac{1}{3} \right) \right] \right\}
e^{4\sigma} \varphi^4  \left( \ln \frac{ e^{2\sigma}
\varphi^2}{m^2} - \frac{25}{6} \right) \nn \\
&&- \frac{1}{4(4\pi)^2} \left\{ f \left( \xi - \frac{1}{6} \right)
+\lambda \xi \left[ 8 \xi + \frac{5}{6} + \frac{10}{3} u +
\frac{1}{u} \left( -3\xi^2 +6\xi +\frac{13}{12} \right)
\right] \right\} \nn \\ && \hspace{5cm} \times  e^{2\sigma} \wh{R}
\varphi^2 \left( \ln \frac{ e^{2\sigma} \varphi^2}{m^2} -3 \right).
\label{18}
\eea
Here $m^2$ is a dimensional parameter. Using this potential one can
estimate the possibility of inducing Einstein's gravity as a result
of a curvature-induced phase transition with order parameter
$e^{2\sigma} \varphi^2$. The result is that such possibility exists
indeed for a variety of theory parameters. Notice also that, in
principle, one can generalize the above potential and calculate it
for the theory (\ref{1}) with the Einstenian part of the action, and
also develop an approach where $e^{2\sigma} > \varphi^2$ and find
the effective potential for the conformal factor in this approach.

As a final remark let us note that the recently proposed average
effective action \cite{21} seems to be a most natural one, to be
used for the study of the effective potential in the infrared
sector. It would be very interesting (albeit not so easy) to try to
apply this action to the theory of quantum gravity, in particular
to the conformal-factor sector (see also \cite{11}).
\vspace{5mm}

\noindent{\large \bf Acknowledgments}

SDO would like to thank I. Antoniadis and R. Percacci for
helpful discussions, and the members of the Dept. ECM, Barcelona
University, for their kind hospitality.
This work has been  supported by DGICYT (Spain), project No.
PB90-0022, by CIRIT (Generalitat de Catalunya), and by the ISF Project
RI1000 (Russia).

\end{document}